\documentclass[12pt,preprint]{aastex}
\def\psr{PSR~J1911$-$5958A }
\def\com{COM~J1911$-$5958A }

\def\ltsima{$\; \buildrel < \over \sim\;$}
\def\gtsima{$\; \buildrel > \over \sim\;$}
\def\lsim{\lower.5ex\hbox{\ltsima}}
\def\gsim{\lower.5ex\hbox{\gtsima}}
\def\lapp{\ifmmode\stackrel{<}{_{\sim}}\else$\stackrel{<}{_{\sim}}$\fi}
\def\gapp{\ifmmode\stackrel{>}{_{\sim}}\else$\stackrel{<}{_{\sim}}$\fi}

\shorttitle{Phase resolved photometry and spectroscopy of COM~J1911$-$5958A}
\shortauthors{Cocozza et al.} 
\begin{document} 

\title{The puzzling properties of the Helium White Dwarf orbiting the
millisecond pulsar PSR J1911$-$5958A in NGC 6752\footnotemark[1]}
\footnotetext[1]{Based on
observations collected at the ESO-Very Large Telescope (Cerro Paranal
Chile), program 071.D-0232A , 073.D-0067A and on data retrived from
ESO Science Archive Facility}

\author{G. Cocozza\altaffilmark{2,3},F. R. Ferraro\altaffilmark{2},
A. Possenti\altaffilmark{4},
N. D'Amico\altaffilmark{4,5}}
\footnotetext[2]{Dipartimento di Astronomia Universit\`a di
Bologna, via Ranzani 1, I--40127 Bologna, Italy;
gabriele.cocozza@studio.unibo.it;francesco.ferraro3@unibo.it}
\footnotetext[3]{INAF--Osservatorio Astronomico di Bologna, via Ranzani 1, I--40127
Bologna, Italy}
\footnotetext[4]{INAF--Osservatorio Astronomico di Cagliari, Loc. Poggio dei Pini,
Strada 54, I--09012 Capoterra, Italy; possenti@ca.astro.it}
\footnotetext[5]{Dipartimento di Fisica Universit\`a di
Cagliari, Cittadella Universitaria, I-09042 Monserrato, Italy;
damico@ca.astro.it}
\medskip

\begin{abstract}
We have used phase-resolved high-resolution images and low resolution
spectra taken at the ESO Very Large Telescope, to study the properties
of the low-mass Helium White Dwarf companion to the millisecond pulsar
\psr (hereafter COM~J1911$-$5958A), in the halo of the Galactic
Globular Cluster NGC 6752. The radial velocity curve confirms that
\com is orbiting the pulsar and allows to derive a systemic velocity
of the binary system nicely in agreement with that of NGC 6752. This
strongly indicates that the system is a member of the cluster, despite
its very offset position ($\sim 74$ core radii) with respect to the
core. Constraints on the orbital inclination ($\gapp 70^\circ$) and
pulsar mass ($1.2-1.5~{\rm M_\odot}$) are derived from the mass ratio
$M_{PSR}/M_{COM}= 7.49\pm0.64$ and photometric properties of
COM~J1911$-$5958A.  The light curve in B-band shows two phases of
unequal brightening ($\Delta$mag$\sim 0.3$ and $0.2$, respectively)
located close to quadratures and superimposed on an almost steady
baseline emission: this feature is quite surprising and needs to be
further investigated.
\end{abstract}

\section{Introduction}
\label{intro}

PSR~J1911$-$5958A is a binary millisecond pulsar (MSP) discovered on
1999 in the nearby Globular Cluster (GC) NGC6752, during a search
performed with the Parkes radio telescope at 1.4 GHz (D'Amico et
al. 2001). Accurate celestial coordinates and orbital parameters for
this binary system have been first derived from $\sim 18$ months of
pulsar timing observations (D'Amico et al. 2002) and recently refined
(Corongiu et al. 2006) using a much longer data span.
The pulsar has a spin period of 3.26 ms and follows an almost circular
orbit ($e\sim 3\times 10^{-6}$, Corongiu et al. 2006)
with an orbital period of 0.84 day.

An unprecedented characteristic of this binary is his position: it is
located at $\sim 6.4\arcmin$ from the cluster center (D'Amico et
al. 2002), corresponding to about 74 core radii (Ferraro et
al. 2003b).  This is the most off--centered pulsar among the entire
catalog of MSPs whose position in the parent cluster is known.  Since
dynamical friction should have driven the binary towards the GC center
in a timescale much shorter than the age of the cluster (Colpi,
Possenti \& Gualandris 2002), the location of \psr has been
interpreted as a evidence for a strong dynamical interaction which
occurred $\lapp 1$ Gyr ago in the cluster core (D'Amico et al. 2001,
Ferraro et al. 2003b), leading to the ejection of the binary towards
the outskirts of NGC 6752. Dynamical events are not unexpected in the
dense stellar enviromnent of a GC, but clear observational signatures
of them are still very rare (e.g. the cases of PSR JB21227+11C in M15,
Phinney \& Sigurdsson 1991, or PSR J0514$-$4002A in NGC 1851, Freire
et al. 2004). Moreover, in the case of \psr it has been conjectured
that a binary black hole of intermediate mass (Colpi, Possenti \&
Gualandris 2002; Colpi, Mapelli \& Possenti 2003) may have been
involved in the interaction.

Stimulated by the implications of the peculiar position of \psr in NGC
6752, we have undertaken a program of optical observations, in the aim
of shedding light on the origin and evolution of the binary and for
assessing its belonging to the cluster.  As a first step, we have
identified the radio pulsar's optical companion (hereafter COM
J1911$-$5958A) by using high-resolution images taken at ESO Very Large
Telescope (Ferraro et al. 2003a; analogous identification was
independently achieved by Bassa et al. 2003). The object turned out to
be a relative faint ($B\simeq22.1$ mag), blue star ($(B-V)=0.1$ mag),
whose position in the color-magnitude diagram lies between the
Main Sequence and the Carbon Oxygen White Dwarf cooling
sequence.  The detailed comparison with theoretical evolutionary
tracks (Serenelli et al. 2002) suggested that it is a Helium White
Dwarf (He-WD) of mass in the range $0.17-0.20~M_{\odot}$, with a
temperature $T_{\rm eff}\approx 11,000~K$, a gravity $\log g\approx
6.2$, a luminosity $L=0.03-0.04~L_{\odot}$, a radius $R=3-4\times 10^9$ cm
and a cooling age in the range $1.2-2.8$ Gyr (Ferraro et al. 2003a).

In this letter we present the results of a second group of photometric
and spectroscopic observations performed at ESO Very Large Telescope
(VLT) in order to determine the radial velocity (\S 2) and light curve
(\S 4) of COM J1911$-$5958A. The mass ratio of the system is derived
in \S 3, as well as contraints on the pulsar mass and the orbital
inclination of the binary.
 
\section{Radial velocity curve of \com}
\label{velcurve}

The analyzed spectra has been retrieved from the ESO Science Archive
Facility. Observations were performed by using the {\it FOcal Reducer
and low resolution Spectrographer 1} (FORS1) mounted at the {\it Antu}
8m-telescope (UT1) of the ESO-VLT on Cerro Paranal (Chile). The
spectra series were obtained during 6 different nights from 2004 July
to 2004 August, using the Long Slit Spectroscopy (LSS) operation mode
and the B-Bessel Grism, which cover the spectral range $3400-5900{\rm
\AA}$. The $1\arcsec$ wide and $6\arcmin.8$ long slit was adopted,
yelding a dispersion of 1.2 ${\rm \AA}$/pixel. The exposure time for
each spectrum ($2500$ sec, $\sim 4\%$ of the system orbital period),
and the total number of different spectra analyzed (23), ensured a
typical signal-to-noise ratio $S/N\simeq 10$ per pixel (measured at
the continuum level), allowing to collect a good phase-resolved set of
data. The data reduction has been perfomed by using the
standard IRAF\footnote{IRAF is distribuited by the National Optical
Observatory which is operated by the Association of Universities for
Research in Astronomy Inc., under cooperative agreement with the
National science Foundation} tools. Raw images has been first
corrected for bias and flat-field and decontaminated from cosmic rays,
then the spectra were extracted and wavelenght calibrated. Using the
task IDENTIFY and polynomial fit with a 4-th order function, the
accuracy of the wavelenght calibration results $\leq 0.01~{\rm \AA}$. Each
spectrum has been corrected for the Earth motion and reported to a
common heliocentric system using the task RVCORRECT.

\begin{figure}[t!]  
\plotone{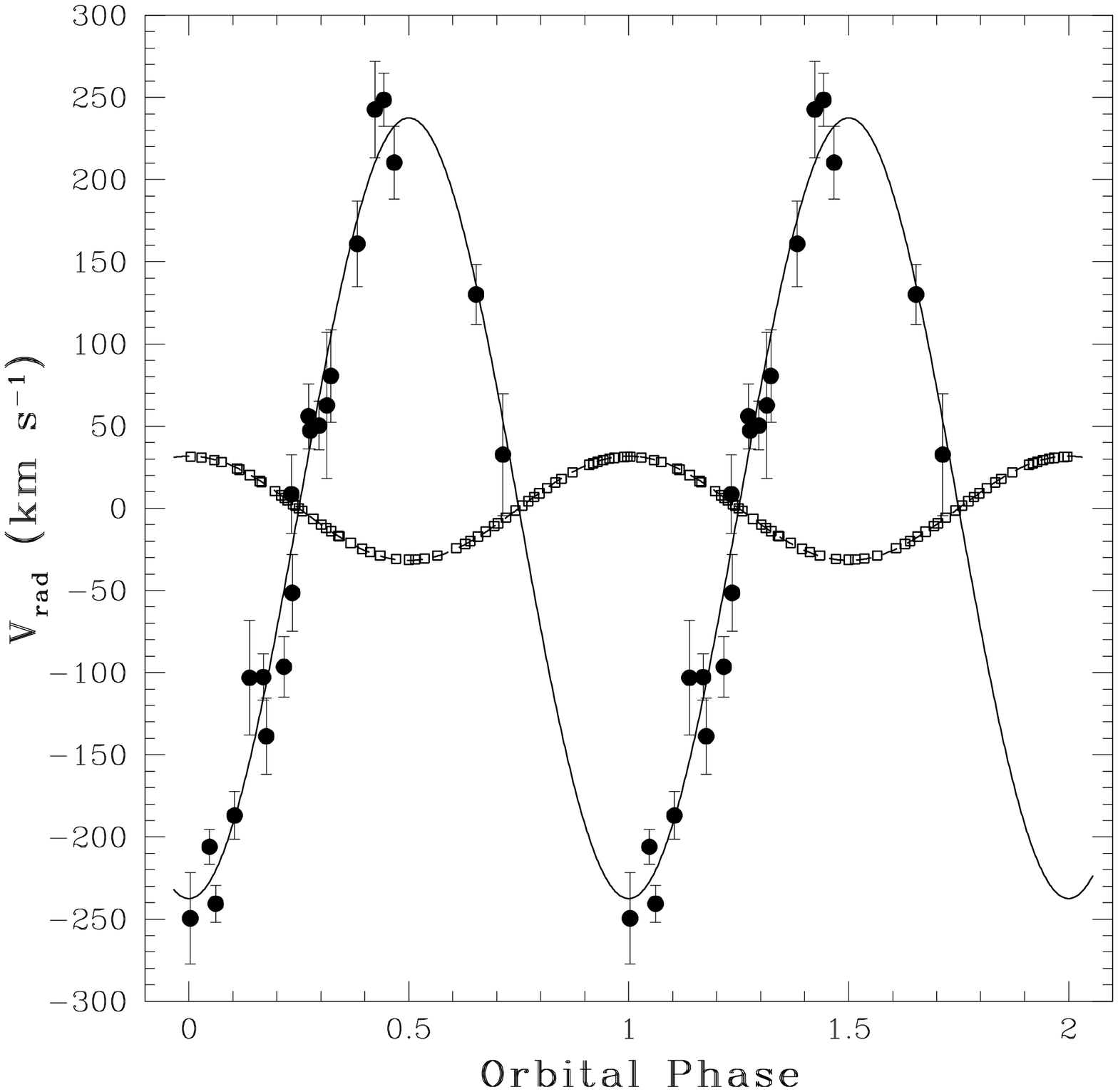}  
\caption {\label{RVel} Velocity curves of COM J1911$-$5958A ({\it large
dots}) and \psr ({\it empty squares}). The data for the pulsar are
derived from timing measurements and the radio ephemeris (Corongiu et
al. 2006). Error bars for the pulsar radial velocity
are smaller than the size of the symbols. The {\it dashed} and the
{\it solid lines} represent the best-fit to the velocity data with a
sinusoidal curve, for the pulsar and the companion star, respectively.
The systemic velocity $\gamma = -28.1$ km s$^{-1}$ of the binary has
been subtracted.}
\end{figure}

The spectral range covered by the spectra allows the observation of
several spectral features, in particular the Hydrogen Balmer-lines
from $H_\beta$ to $H(7).$ The Doppler-shifted wavelenght of each
line has been measured by using the SPLOT task, fitting with a
gaussian function all the features. The wavelenght of each line has
been converted to a radial velocity (RV), then all the RVs measured for
each spectra has been averaged and a mean value was obtained
(accounting for the heliocentric correction).  Only the 21 RV
determinations resulting from the average of at least five measures
have been used.  The mid-point of each observations has been converted
to a orbital phase by adopting the orbital period and the epoch of the
ascending node for the pulsar orbit as given by the radio ephemeris
(Corongiu et al. 2006). Phases 0.0 and 0.5 correspond
to the quadratures, phase 0.25 to the inferior conjunction of the
companion (i.e. when \com is at its closest position with respect to
the observer) and phase 0.75 to the superior conjunction.  In order to
determine the amplitude $K$ of the radial velocity curve and the
systemic velocity $\gamma$ of the binary system, we have fitted the
data by using a function which is the sum of a constant and a
sinusoid, adequate to describe the almost perfectly circular orbit of
COM J1911$-$5958A.  The best fit curve yields $K=237.5\pm 20.0$ km
s$^{-1}$ and $\gamma = -28.1 \pm 4.9$ ($1\sigma$ uncertainties are
used here and everywhere in the paper) and is reported in Figure
\ref{RVel}.

\section{The mass ratio}

The systemic velocity of the binary system is in agreement with the
published radial motion of the globular cluster ($v_{\rm NGC6752} =
-27.9 \pm 0.8$, Harris et al. 1996, catalog revision 2003). This lends
further support to the cluster membership of the binary. Given the
central 1-D dispersion velocity of NGC 6752 (9-15 km s$^{-1}$, Drukier
et al. 2003), the expected 1-D dispersion velocity for objects of mass
$1.4-1.7$ M$_{\odot}$ (corresponding to the most likely total mass of
the binary, see later) and located at the projected position of \psr
with respect to the cluster center is 2-3 km s$^{-1}$ (Mapelli,
private communication). This is fully compatible with the value of the
difference $\Delta v_{\rm 1D}=|{\gamma}-v_{\rm NGC6752}|\lapp 6$ km
s$^{-1}.$ Moreover, the small value of $\Delta v_{\rm 1D}$ may
indicate that the binary is now near apoastron of a highly elliptical
orbit in the potential well of the globular cluster.  In fact, were
the binary on an almost circular orbit at 74 core radii from the GC
center, its line of sight velocity with respect to the cluster center
(Sabbi et al. 2004), as estimated from the enclosed mass, would be of the
order $\gapp 12$ km s$^{-1}.$ All these considerations support the
hypothesis that the binary has been recently kicked out of the core of
NGC 6752 due to a dynamical interaction (see \S 1).

We are in the position of inferring the ratio between the masses of
the two stars in the binary.  The mass function of the pulsar, as
measured from radio observation, is (Corongiu et al. 2006):
\begin{equation}
f(M_{PSR}) = {{M_{\rm
COM}^{3} \sin^{3}i}\over{(M_{\rm COM}+M_{\rm PSR})^{2}}} = 
0.002687849(6)~{\rm M_{\odot}}, 
\label{eq:mfpsr}
\end{equation} 
\noindent whereas the mass function of the companion, derived from the present
spectroscopic observations results:
\begin{equation}
f(M_{\rm COM}) = {{M_{\rm
PSR}^{3} \sin^{3}i}\over {(M_{\rm COM}+M_{\rm PSR})^{2}}} 
= {{K^3 P_{\rm orb}}\over{2\pi G}}=\\
1.13\pm0.29~{\rm M_{\odot}},
\label{eq:mfcom}
\end{equation} 
\noindent where $P_{\rm orb}=0.83711347700(1)$ days (Corongiu et
al. 2006), K is the amplitude of the radial velocity curve,
$M_{\rm PSR}$ and $M_{\rm COM}$ are the masses of the pulsar and the
companion, $i$ is the inclination of the normal to the orbital plane
with respect to the line-of-sight, and the brakets report the
$1\sigma$ errors on the last significant digit. The mass ratio
$q=M_{\rm PSR}/M_{\rm COM}$ can be derived by combining
eq. (\ref{eq:mfpsr}) with eq. (\ref{eq:mfcom}), and results $q=7.49\pm
0.64.$

\begin{figure}[t!]   
\plotone{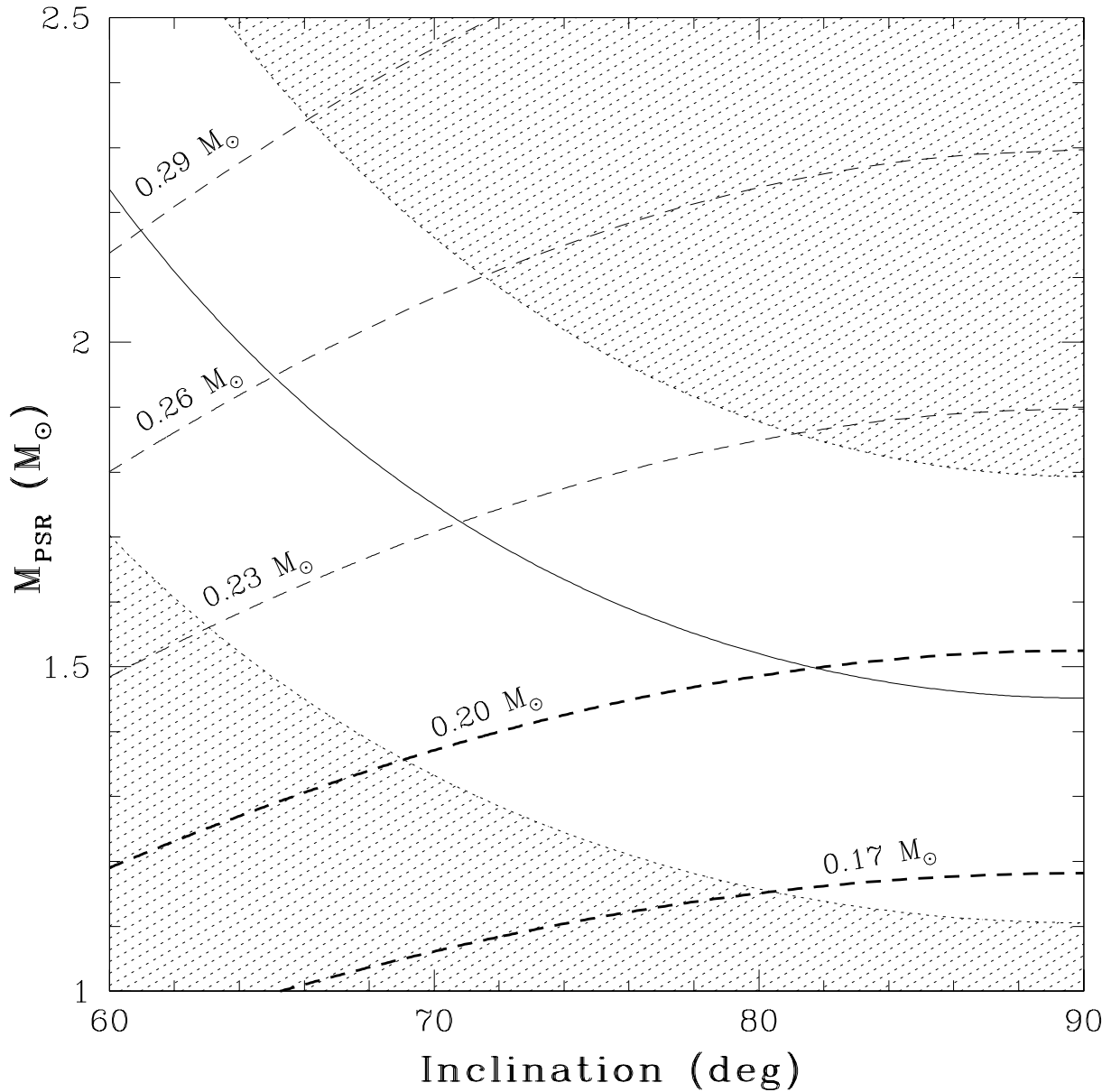}
\caption{\label{MvsI}
Mass of \psr and orbital inclination of the binary. The allowed range
of values are constrained to lie within the strip whose borders ({\it
dotted lines}) are the $1\sigma$ boundaries derived from the mass
ratio of the system. Lines of constant mass for \com are also shown
({\it dashed lines}) and labeled with the assumed mass value. If \com
is a He-WD, the space of parameters is additionally constrained (see
Ferraro et al. 2003a) between the lines corresponding to companion
masses 0.17 and 0.20 ${\rm M_\odot}$ ({\it thick dashed lines}).}
\end{figure}

Solving separately for the masses of the two stars would require a
determination of the orbital inclination: the relation between $M_{\rm
PSR}$ and $i$ for different values of $M_{\rm COM}$ is displayed in
Figure \ref{MvsI}, for a reasonable choice of the neutron star mass
($1-2.5~{\rm M_\odot}$, Shapiro \& Teukolsky 1983). The measured mass
ratio (with its uncertainty) selects a narrow strip of allowed
parameters in Figure \ref{MvsI}. In particular, $M_{\rm PSR}>1.1~{\rm
M_\odot}$ and $M_{\rm COM}>0.16~{\rm M_\odot}$ ($1\sigma$
limits). Moreover, $M_{\rm COM}\lapp 0.30~{\rm M_\odot}$ and the
orbital inclination is $\gapp 60^\circ.$ Tighter constraints can be
obtained in the hypothesis that the companion is a He-WD: in this
case, the range of masses determined by Ferraro et al. (2003a) implies
that the system is almost edge-on ($i\gapp 70^\circ$) and the pulsar
mass is in the range $1.2-1.5~{\rm M_\odot}.$ It is worthwhile to note
that the latter mass interval brackets the values of all neutron star
masses accurately measured so far (Lorimer 2005), but one case (Nice et
al. 2005).   

\section{The puzzling light curve of \com}
\label{lightcurve}

The photometric observations were performed in service mode at the
{\it Antu} 8m-telescope of the ESO-VLT during two nights in 2003,
March and May (ESO Program ID 071.D-0232), and six nights in 2004,
August (ESO Program ID 073.D-0067A). All the images were acquired
using B-band filter in high-resolution mode with the FORS1 camera. In
this configuration the instrumental pixels scale is $0.1\arcsec$
pixels$^{-1}$ and the Field of View of the $2048\times2048$ pixels
Tektronix-CCD is $3\arcmin.4\times3\arcmin.4$.  The data comprise
twenty-one 600s and five 360s exposures, centered roughly less then
$1\arcmin$ far from the {\psr} nominal position (D'Amico et al. 2002).
The observations on 2004 were planned in order to evenly distribute
along the orbit of the binary system.

From all the original frames we have extracted a subimage of
$500\times 500$ pixels, roughly centered on the nominal position of
PSR J1911$-$5958A. The scientific images have been reduced using
ROMAFOT, a package specifically developed to achieve accurate
photometry in crowded fields (Buonanno et al. 1983); it enables the
visual inspection of the quality of point-spread function (PSF)
procedure. PSF best-fitting has been performed for all the images
separately, and the mask with the star position obtained from the
best-quality image was adapted to each image. The instrumental
magnitudes have been reported to a common photometric system, then we
have obtained a catalog with the coordinates and the instrumental
magnitudes for all the stars common to all the images.  We have
performed the photometric calibration using two different and
indipendent methods. First, magnitudes have been matched to four
standard stars (Landolt 1992) observed under photometric conditions in
the B-band.  Then, all the $\sim 100$ stars in our catalog in common
with the B-band catalog published by Ferraro et al. (2003) have been
used to derive the photometric zero-point. The two resulting
calibrations are fully consistent within a few hundredths of
magnitude.

\begin{figure}[t!]  
\plotone{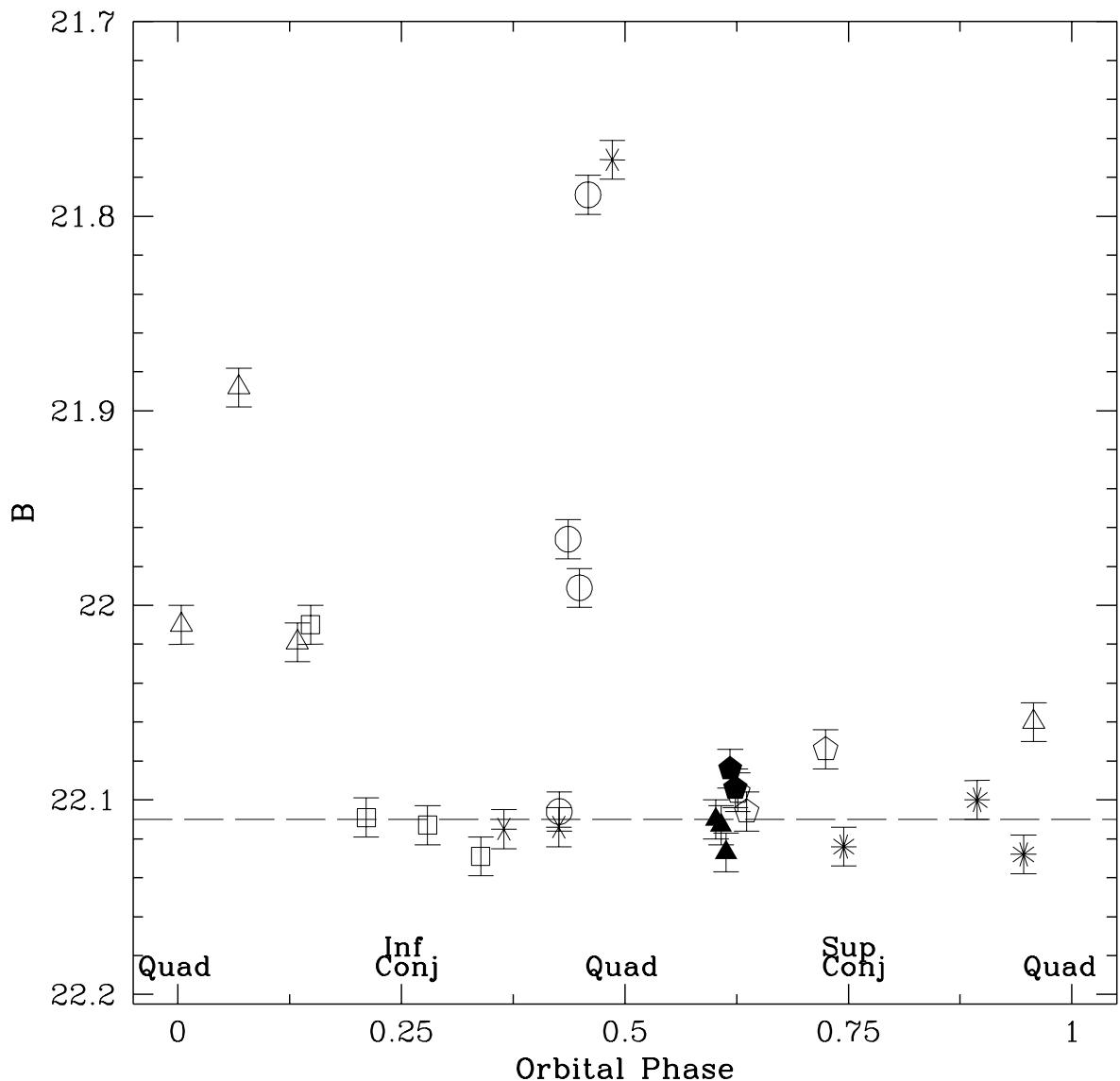}
\caption{\label{LC} Light curve of COM J1911$-$5958A: the {\it filled}
symbols represent observations performed on two different nights on
March and May 2003, whereas the {\it empty} and {\it starred} symbols
represent the data collected on August 2004. Different marks refer to
different nights of observation. The horizontal {\it dashed line}
represents the off-peaks mean B-magnitude of the source, calculated
averaging the data at orbital phases in the ranges (0.2$-$0.4) and
(0.6$-$0.95). The phases of quadrature and conjunction of \com are
reported for clarity.}
\end{figure}

A periodicity search was carried out on the photometric data set using
GRATIS (GRaphycal Analyzer of TIme Series), a software package
developed at the Bologna Astronomical Observatory (see previous
applications in Ferraro et al. 2001).  Since the period ($P=0.839\pm
0.002$ day) obtained from the \com variability curve turned out to be
fully consistent with that obtained from the radio time series of PSR
J1911$-$5958A, a phased light curve has been produced (Figure
\ref{LC}) assuming the same orbital parameters used in \S 2.  The
different symbols in Figure \ref{LC} mark series of B-magnitudes
obtained in different observing nights (hence different companion's
orbits): in particular, empty and starred symbols are used for the
pointings of 2004, whereas filled symbols (clustering at about phase 0.6)
refer to observations of 2003.

The result is really surprising: \com shows two phases of strong
enhancement of the luminosity located close to the quadratures and
extending for about $20\%$ of the orbit each. The primary maximum
occurs between phases 0.5 and 0.6 (a more precise positioning is
prevented by the lacking of useful data in that range) with a flux
variation of $\gapp 0.3$ mag in less than $1$ hr: orbital location and
rapidity of the brightening are confirmed in two different orbits. A
secondary maximum, corresponding to a brightening of $\sim 0.2$ mag
appears at about phase 0.05, presenting a less steep raise with
respect to the primary peak (in this case only the decay from the
maximum is observed in two different orbits). Outside the peaks, the
source displays an almost steady luminosity at an average $B=22.11\pm
0.02.$ In particular the B-band luminosity at orbital phase $\sim 0.6$
does not show any significant fluctuation between the 2003 and the
2004 data (the pointings of 2003 unfortunately did not cover other
orbital phases).

These features are unusual and intriguing.  The modulation of the
optical light curve of the companion to a MSP is most often driven by
the heating of one emisphere of the star (whose rotation is
syncronized with the orbital period) due to the radiation coming from
the pulsar: see for example the cases of PSR B1957+20 (Callanan, van
Paradijs, \& Rengelink 1995) and PSR J2051-0827 (Stappers et al. 2001)
in the Galactic field and 47 Tuc U and 47 Tuc W in the GC 47 Tucanae
(Edmonds et al. 2001, 2002). For \psr the square of the ratio
($\xi_1\sim 100$) between the orbital separation ($\sim 4.5-5 {\rm
R_\odot}$) and the radius of the companion (inferred from the
off-peaks luminosity and the effective temperature, Ferraro et
al. 2003a) is much larger than the ratio ($\xi_2\sim 25$) between the
rotational power of the pulsar\footnote{It is calculated with the
usual formula $\dot{L}_{\rm rot}=(4\pi^2)I\dot{P}/P^3$ where $I$
(assumed equal to $10^{45}$ g cm$^2$) is the neutron star moment of
inertia, $P$ the spin period and $\dot{P}$ the spin period
derivative. Given the observed proper motion and the off-centered
position of \psr in NGC~6752, the value of $\dot{L}_{\rm rot}$ can be
underestimated at most a factor $3$ as a consequence of the centrifugal
acceleration of the pulsar and of the combined effects of the
gravitational potentials of the Galaxy and of the globular cluster
(Corongiu et al. 2006).  This possible correction
leaves the discussion about the heating effect unchanged.} and
the bolometric off-peaks luminosity of COM J1911$-$5958A. This implies
that the modulation due to the heating effect ($\Delta{\rm
(mag)}=2.5\log[1+(\eta/2)(\xi_2/\xi_1^2)]$, where $\eta<1$ is the
efficiency of the process) should be negligible ($\Delta{\rm
(mag)}\lapp 0.001$) and in fact no flux enhancement is detected around
phase 0.75, when the side of the companion facing the pulsar is
visible.

Ellipsoidal variations due to the tidal deformation of the companion
are known to produce a light curve with two peaks at quadratures (see
for example the case of PSR J1740$-$5340 in NGC 6397, Ferraro et
al. 2001), but the light curve is expected to have maxima of equal
amplitude and clear minima (of unequal depth) at conjunctions.
Moreover, tidal deformations are expected to be insignificant for a
companion whose radius is $\sim 20$ times smaller than the radius of
its Roche lobe.

Accretion of matter onto a compact object can generate a variety of
modulated optical emission (e.g. Frank, King \& Raine 2002), but
neither the neutron star nor the He-WD in this system can be suitable
sources of plasma feeding accretion-related processes at the present
epoch. The timing stability and the extension of the time span of the
radio observations of \psr (Corongiu et al. 2006) tend
also to exclude the existence of a residual accretion disk around the
pulsar or the presence of a third optically faint body which is now
pouring mass in the binary.

One may wonder if the optical modulation can be intrinsic to the
He-WD. Non-radial pulsations of WDs can produce optical fluctuations
at a level of $\sim 0.2$ mag, but for a star with $T_{\rm eff}\approx
11,000~K$ the expected modulation occurs at periods significantly
shorter than 0.84 days (e.g. Bergeron et al. 2004). However, a few
high magnetic field ($\sim 10^8$ G) isolated WDs (see e.g. PG
1031+234, Piirola \& Reiz 1992, and EUVE J0317$-$855, Barstow et
al. 1995) display variations of $\lapp 0.3$ mag at the supposed
rotational period of the star. In the framework of an oblique rotator
model for the WD, these photometric modulations have been interpreted
with changes with the rotational phase of the mean magnetic field
strenght over the visible stellar surface (which in turn affects the
opacity along the line of sight: Ferrario et al. 1997).  A suitable
geometry (leading to an alternate exposure of both the magnetic polar
caps of the WD) may in principle produce a double peaked light curve,
but we note that all the aforementioned effects have been observed
only in relatively massive WDs ($\sim 0.5-1.0~{\rm M_\odot}$) insofar.

Given this puzzling scenario, further data are clearly necessary for
constraining the origin of the optical modulation in this system, namely
phase resolved multi color (UBVRI) photometry, complemented with
higher resolution spectroscopy.
In this respect, polarimetric observations would be 
particularly enlightening, since the detection of 
linear/circular polarization variation  at the 
rotational/orbital period would be a strong signature of a highly magnized
WD.

\acknowledgements

We thank E. Sabbi for assistance with the Phase II preparation
procedures and P. Montegriffo for assistance with the use of the
GRATIS software.  Financial support to this reasearch has been
provided by the Agenzia Spaziale Italiana (ASI) and by the {\it
Ministero dell'Istruzione, dell'Universit\`a e della Ricerca} (MIUR).


\clearpage 

\clearpage

\clearpage

\end{document}